\documentstyle[amssymb,epsfig,preprint,aps,prb]{revtex}
\tightenlines
\newcommand{\symcircle}{\raisebox{-1pt}{\large $\circ$}}
\newcommand{\symsquare}{\raisebox{-1pt}{\scriptsize $\Box$}}
\newcommand{\symdiamond}{\raisebox{-1pt}{\scriptsize $\Diamond$}}
\newcommand{\symtriangleu}{\raisebox{-0.5pt}{\tiny $\bigtriangleup$}}
\newcommand{\symtriangled}{\raisebox{1pt}{\tiny $\bigtriangledown$}}

\begin{document} 

\title{Do cylinders exhibit a cubatic phase?}
\author{Ronald Blaak, Daan Frenkel and Bela M. Mulder}
\address{FOM Institute for Atomic and Molecular Physics, Kruislaan 407,\\
1098 SJ Amsterdam, The Netherlands.} 
\date{\today}
\maketitle

\begin{abstract}
We investigate the possibility that freely rotating cylinders with an
aspect ratio $L/D=0.9$ exhibit a cubatic phase similar to the one
found for a system of cut-spheres. We present theoretical arguments
why a cubatic phase might occur in this particular system. Monte Carlo
simulations do not confirm the existence of a cubatic phase for
cylinders. However, they do reveal an unexpected phase behavior
between the isotropic and crystalline phase. 
\end{abstract}

\section{Introduction}
\label{sec:intro}

The phase behavior of apparently simple systems can be surprisingly
rich. In 1957 Alder and Wainwright reported the first simulations of
a system of hard spheres \cite{Alder:1957JCP}. They showed that, upon
increasing density, this system  exhibits a phase transition from an
homogeneous liquid to a crystalline FCC phase, even though there is no
attractive interaction.

An almost equally simple system that exhibits an even richer phase
behavior, is the hard spherocylinder model, that was introduced by
Onsager in 1949 \cite{Onsager:1949AAS}. Onsager used this model to
show that hard, elongated, rodlike particles must undergo a transition
from the isotropic to nematic phase. Subsequently, computer
simulations, revealed that a smectic A and two different crystalline
phases can exist
\cite{Stroobants:1986PRL,Bolhuis:1997JCP1,Frenkel:1988NAT},  depending
on the aspect ratio. Recent extensions of the Onsager theory to higher
densities and smaller aspect ratios can account for these novel phases
\cite{Somoza:1990PRA,Poniewierski:1990PRA,Graf:1997PCPS,Graf:1999PRE}.  

In order to study the behavior of disklike objects Eppenga and Frenkel
performed simulations of a system of infinitely thin disks
\cite{Eppenga:1984MP}. By construction, the packing fraction of such a
system is zero. As a consequence, after aligning the particles, the
system can easily be compressed, nevertheless, the nematic to columnar
transition for this system has been studied \cite{Bates:1998PRE}.
A different model for disklike object was proposed by Veerman and
Frenkel \cite{Veerman:1992PRA}. They used  cut-spheres, which are
obtained by symmetrical slicing off the polar caps of a sphere. As
a function of the length over diameter ratio of these particles, they
connect the infinitely thin discs with the spheres. Surprisingly,
cut-spheres with an aspect ratio of about 0.2, appear to exhibit an
orientationally ordered phase with a cubic symmetry, which is called the
cubatic phase, even though these particles themselves have  a
cylindrical symmetry.  
   
In this article we investigate a possible explanation of the existence
of the cubatic phase for cut-spheres. To this end we study, by
computer simulation, a system of freely rotating cylinders. In Sec.
\ref{sec:cubatic} we present a theoretical basis for looking at
cylinders. In Sec. \ref{sec:simulation} we elucidate some of the
techniques used in our Monte Carlo simulation. This involves the
overlap criterion of these particles and an additional type of move
that we used in our simulations. In Sec. \ref{sec:result} we show
the main results from the computer simulations and discuss
the unexpected behavior of cylinders between the isotropic
and crystalline phase. The symmetry of this new phase is analyzed in
Sec. \ref{sec:symmetry}. In Sec. \ref{sec:free} we determine the
coexistence between the isotropic liquid and the crystalline phase by
means of a free energy calculation. In
Sec. \ref{sec:discussion} we discuss the main results and give
some suggestions for future research.

\section{Cubatic phase}
\label{sec:cubatic}

The cubatic phase is a long-range orientationally ordered phase without
any positional order of the particles. In a uniaxial nematic phase the
particles, e.g. rods, tend to align along a single preferred
direction. In a biaxial phase for single component systems,
e.g. biaxial ellipsoids \cite{Camp:1997JCP1}, the different molecular
axes of the particles align in three different orthogonal
directions. In the special case that these three directions are also
equivalent, in the sense that each molecular axis is with equal
probability in any of the three directions, we obtain an
orientationally ordered phase with cubic symmetry. In the absence of 
translational order this is the elusive cubatic phase.

Evidence for this cubatic phase was first found in computer
simulations of cut-spheres with an aspect ratio $L/D=0.2$
\cite{Veerman:1992PRA}. However, subsequent theoretical work indicates
that particles, consisting of three perpendicular, elongated rods of
approximately the same length, at least in the Onsager limit of
infinite aspect ratios, should also form this cubatic phase 
\cite{Frenkel:1991,Blaak:1998PRE}. Experimentally, however, this
cubatic phase has never been observed.  

\begin{center}
\begin{figure}[h]
\epsfig{figure=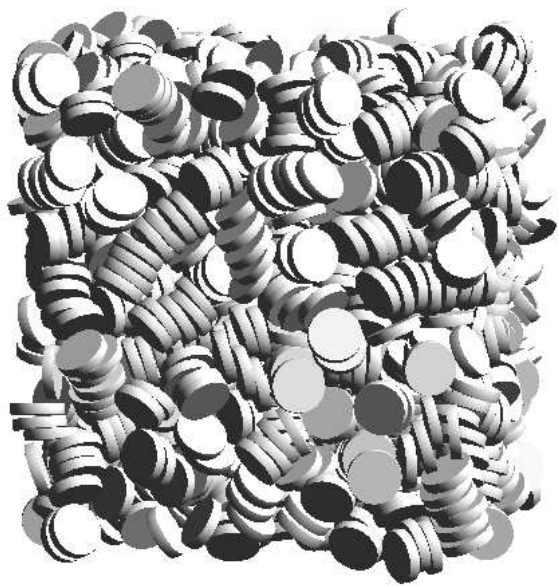,width=8cm}
\vspace{4cm}
\caption[a]{\label{fig:cubatic} R. Blaak, Journal of Chemical Physics}
\end{figure}
\end{center}

Figure \ref{fig:cubatic} shows a snapshot of the cubatic phase of
cut-spheres. The figure suggests that the disklike objects tend to
form stacks of several particles in an approximately cylindrical
shape. These cylindrical clusters are then arranged in such a way that
there is an overall cubatic order. As a first attempt to explain the
cubatic phase for cut-spheres we will attempt to describe the system
on the level of these aggregates. This is a simplification, because the
stacks in Fig. \ref{fig:cubatic} are obviously not perfect cylinders
but polydisperse in both length and shape. Yet, in what follows, we model
this system as a collection of monodisperse, hard cylinders.

In order to select the optimal aspect ratio that might give rise to a
cubatic phase, we first consider the excluded volume of two
cylinders. Onsager already gave an expression for the excluded volume
${\cal E}$ of two arbitrary cylinders with lengths $L_i$ and diameters
$D_i$ \cite{Onsager:1949AAS} 

\begin{eqnarray}
\label{eq:excl-volume}
{\cal E}(\gamma) = & \frac{\pi}{4} D_1 D_2 (D_1+D_2) \sin \gamma +
 L_1 L_2 (D_1 + D_2) \sin \gamma  \nonumber \\ 
 & + L_1 \left( \frac{\pi}{4} D_1^2 + D_1 D_2 E(\sin \gamma) +
 \frac{\pi}{4} D_2^2 | \cos \gamma| \right) \nonumber \\ 
 & + L_2 \left( \frac{\pi}{4} D_2^2 + D_1 D_2 E(\sin \gamma) +
 \frac{\pi}{4} D_1^2 | \cos \gamma| \right) 
\end{eqnarray}
where $\gamma$ is the angle between the two main axes of the cylinders
and $E(x)$ is the complete elliptical integral of the second kind.
 
\begin{center}
\begin{figure}[h]
\epsfig{figure=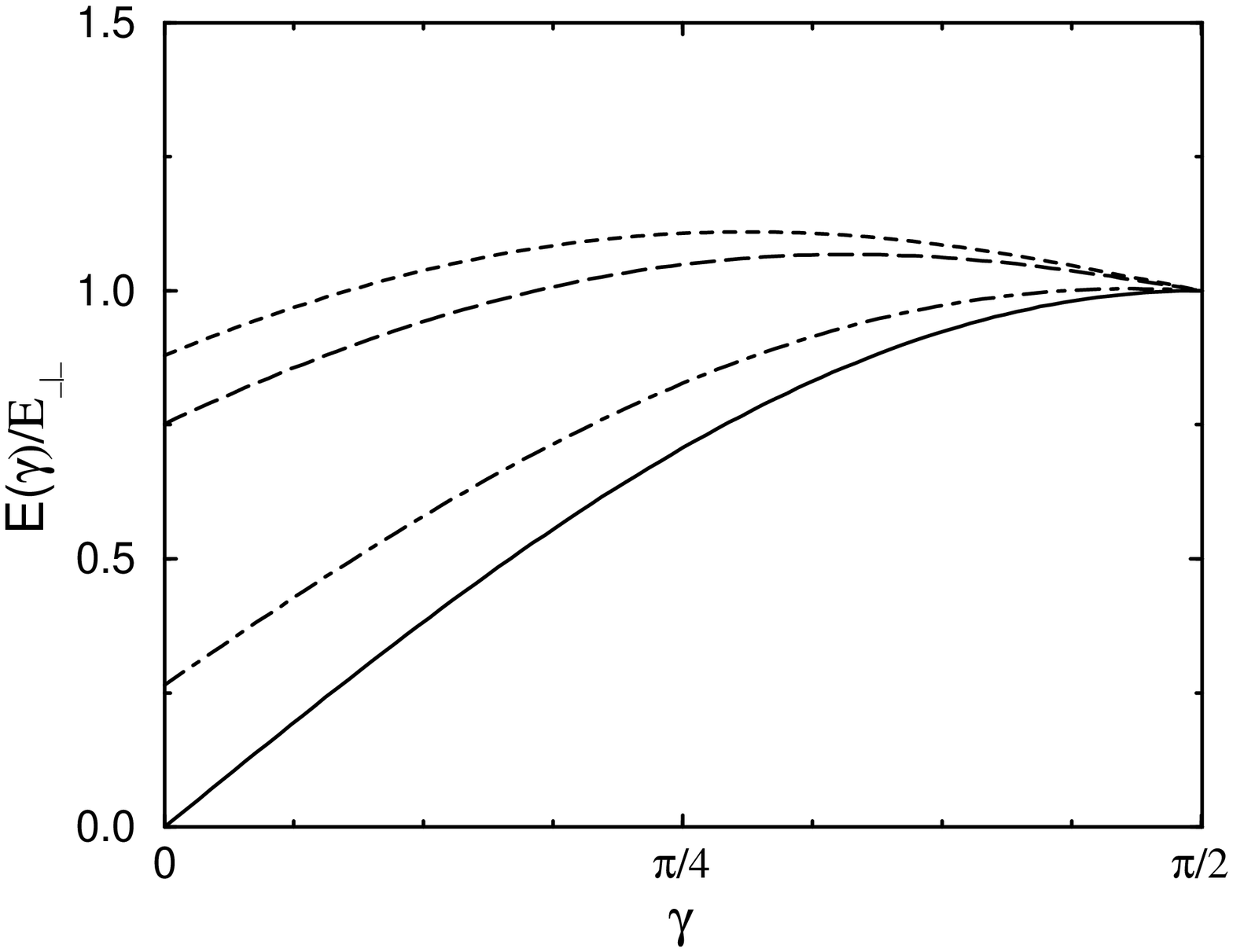,width=8.5cm}
\vspace{4cm}
\caption[a]{\label{fig:excl_volume} R. Blaak, Journal of Chemical Physics}
\end{figure}
\end{center}

In Fig. \ref{fig:excl_volume} we have plotted the excluded volume of
two identical cylinders as function of the angle $\gamma$, for several
aspect ratios. To facilitate comparison we normalized the excluded
volume to unity for perpendicular orientations. Note that for the aspect
ratios, zero and infinity, ${\cal E}(\gamma)$ reduces to the same
monotonically increasing function. This reflects the tendency of
strongly anisotropic particles to align. For an aspect ratio of the
order unity, however, we see that the maximum excluded volume is found
between the parallel and perpendicular orientation. In fact, for the
latter two orientations the exclude volume exhibits a minimum, although
the deepest minimum always corresponds to a parallel orientation.

\begin{center}
\begin{figure}[h]
\epsfig{figure=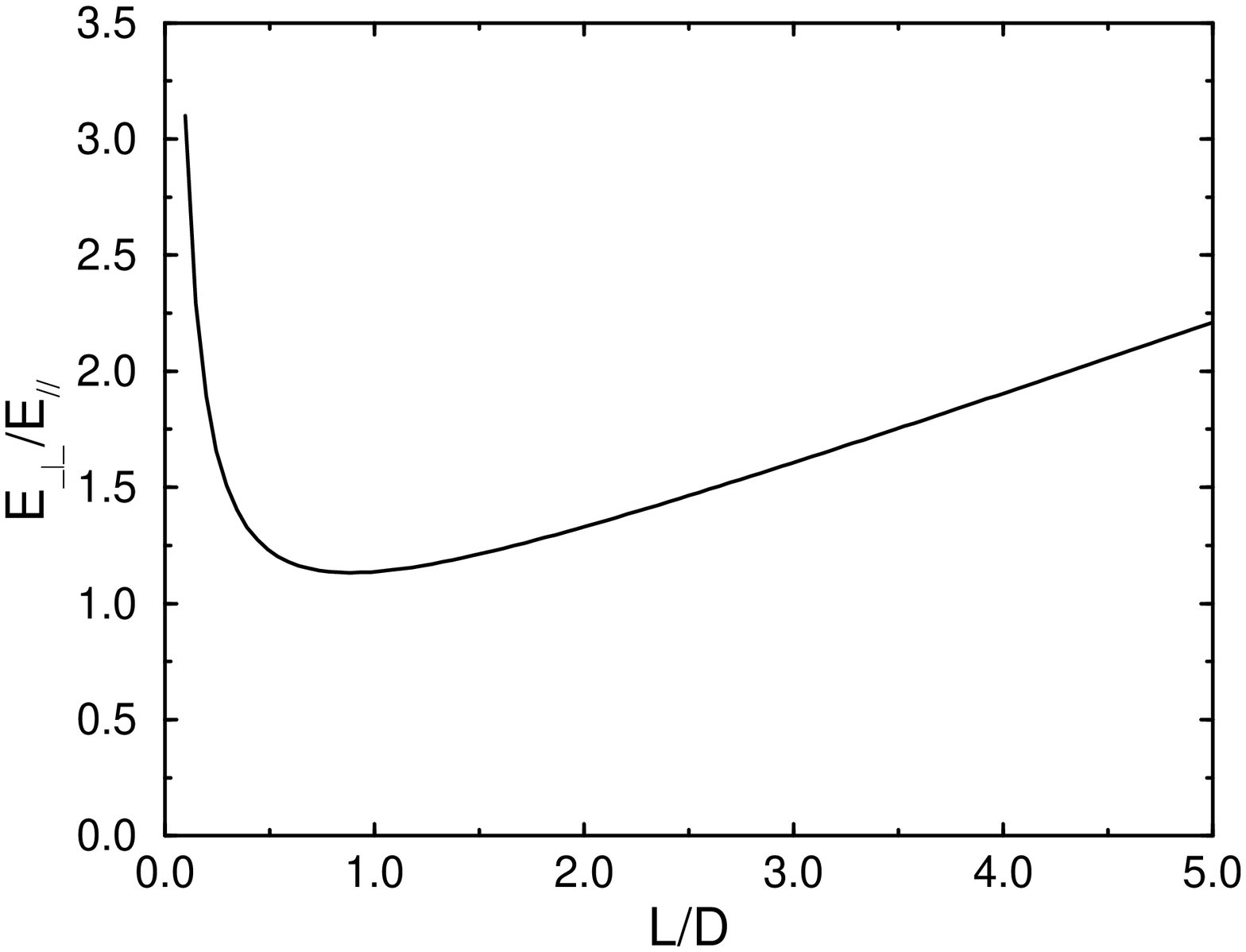,width=8.5cm}
\vspace{4cm}
\caption[a]{\label{fig:excl_ratio} R. Blaak, Journal of Chemical Physics}
\end{figure}
\end{center}

In a cubatic phase both parallel and perpendicular orientations should
be present. However, it is logical to look for that aspect ratio that
minimizes the relative difference between both minima. The fact that
parallel orientations lead to a smaller excluded volume, and are
therefore favored, could be compensated by the fact that forming a
cubatic phase would lead to a higher orientational entropy. This is
because particles in a cubatic phase can be aligned in three different
directions instead of a single one as in the case of only parallel
orientations. In Fig. \ref{fig:excl_ratio} we have plotted the ratio of the 
perpendicular and  parallel orientation as a function of the aspect
ratio. The minimum of the curve occurs for the aspect ratio
$L/D=\sqrt{\pi}/2 \approx 0.886$, where the excluded volume of
perpendicular orientations is only 1.133 times larger than that of
parallel orientations.

\section{Simulation}
\label{sec:simulation}
In our search for a possible cubatic phase for hard cylinders, we
performed Monte Carlo simulations of a system of freely rotating, hard
cylinders with an aspect ratio $L/D=0.9$. 

\subsection{Overlap criterion}
\label{sec:overlap}

In \cite{Allen:1993ACP} an overview is given of the most widely used
overlap criteria for convex hard particles. The one for cylinders is
only outlined. It consists out of three steps.
\begin{enumerate}

\item {\em spherocylinder overlap}: For the first step we replace the 
cylinders by spherocylinders by adding hemispheres on both sides of
the cylinders. We then compute the closest distance between the
spherocylinders. If the spherocylinders do not overlap the cylinders
will not either. If the spherocylinders do overlap we can only
conclude that the cylinders overlap if the vector of shortest distance
intersects both cylindrical parts.

\item {\em disk-disk overlap}: If the last condition is not satisfied
we need to continue with the second step. In this step we check
whether there is an overlap of the  flat faces of the cylinders.

\item {\em disk-cylinder overlap}: If no overlap is found in the
second step, and the spherocylinders in the first step did overlap 
we need to proceed with  the third step, which is the most
time-consuming one. We have to check whether one of the flat faces of a 
cylinder overlaps with the cylindrical hull of the other cylinder. 
\end{enumerate}

According to \cite{Allen:1993ACP} this last problem reduces to calculating
overlaps between two planar ellipses, and hence is a special case of
the overlap between two ellipsoids. It is this last step which is the
most difficult.

For ellipsoids there exist two different criteria to determine
whether there is an overlap or not. The first is due to
Vieillard-Baron \cite{Vieillard:1972JCP} and the second due to Perram and 
Wertheim \cite{Perram:1985JCOP}. Both criteria determine whether there
exists an overlap without calculating any points which both particles
have in common. But although the intersection of the plane in which
the flat face lies with the cylindrical hull of the other particle is
an ellipse, the ellipse is not necessary complete. This happens if the
plane also intersects with one of the flat faces of the second
cylinder. In that case we need to check whether there is an overlap
with the incomplete ellipse.

The alternative route is to look at the problem as the overlap between
a flat face of one cylinder with a sequence of circles of the second
particle. This results in a fourth order polynomial equation and can
in principle be solved analytically. The main problem with this
approach is that during simulations it will result in severe problems,
and not always find the correct outcome. The reason for this is that
the coefficients of the polynomial equation are inversely proportional
to the sine of the angle between the particles. Therefore the
coefficients can differ by orders of magnitudes, which makes it hard
to solve and not always accurate.

The method we applied for the last step in the overlap criterion
therefore is a direct minimization of the distance between a point on
the edge of the first cylinder and a point on the axis of the second
cylinder. Although it is slower, it has proven to be very reliable.

\subsection{Flip-move}
\label{sec:flip-move}
In Monte Carlo simulations we wish to sample the complete
configuration-space in an efficient way. In our Monte Carlo
simulations we keep the number of particles $N$, the pressure $P$ and
the temperature $T$ are fixed. We performed several distinct type of
trial moves. We try to translate and rotate particles by a small
amount. In addition we perform volume changing moves in order to
equilibrate the density under the applied pressure. For details about
these standard techniques, we refer to \cite{Book:Frenkel-Smit}.

At high densities, we can only perform relatively small trial moves,
because otherwise the acceptance of the moves becomes negligibly
small. However, this may create a kinetic barrier to equilibrate a
cubatic phase, in which particles are parallel or perpendicular to
each other, while intermediate orientations are less likely due to
steric hindering. It is difficult to achieve a rotation of a cylinder
over $90^{\circ}$ by a succession of small rotational moves.

In order to sample the configuration space of this system more
efficiently, we introduce an extra trial move: the flip move. Rather
than waiting for the rare reorientation of a cylinder over
$90^{\circ}$ through a succession of small angular jumps, we attempt to
rotate it directly to a perpendicular orientation. 
The flip-move is a proper Monte Carlo move that satisfies detailed
balance. By construction it is such, that it does not break the
symmetry of an isotropic phase. 
In the case of short
cylinders this trial move is surprisingly effective.  

\begin{center}
\begin{figure}[h]
\epsfig{figure=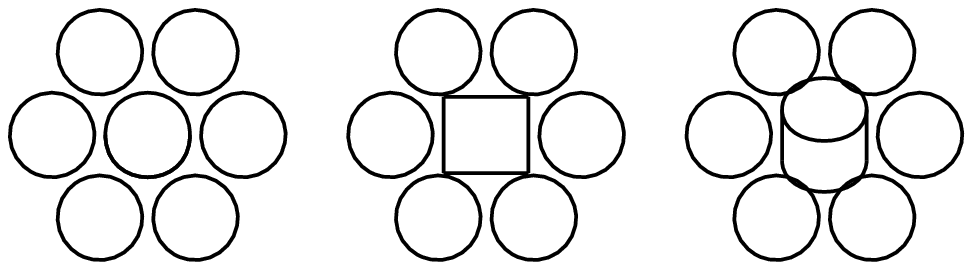,width=8cm}
\vspace{4cm}
\caption[a]{\label{fig:flip} R. Blaak, Journal of Chemical Physics}
\end{figure}
\end{center}

To illustrate this we have shown in Fig. \ref{fig:flip} three steps of
the continuous rotation of a short cylinder in a dense ordered
phase. The first picture shows that for parallel particles there is
still some space to move. However, if the cylinder in the middle is
rotated by  $45^{\circ}$, its dimension in the plane changes
causing it to overlap with neighboring particles. As the rotation
continues the excluded volume of the cylinder decreases 
again. By including the flip-move we bypass the unfavored
intermediate state. 

During our simulations on average 1 out of every 10 rotations will be  
an attempt  to flip a particle. To this end, we select at random an
orientation in the plane perpendicular to the direction of the
particle and accept it if there is no overlap in the new situation. The 
acceptance of this move depends of course strongly on the aspect ratio
and density, but can be as large as 5\%, even in a crystalline phase.

\subsection{Nature of crystalline phase}
\label{sec:crystal}

There are several relevant crystalline structures to study for a
system of cylinders with an aspect ratio of order unity. The highest 
possible density in a system of cylinders is reached when all
cylinders are perfectly aligned in a close-packed structure with
the particles ordered in hexagonal layers. This leads to a maximum 
packing fraction $\phi = \pi/\sqrt{12}$.

In the case of spheres there are many different possible ways to stack
the layers to form a regular crystal. The simplest stackings are the
face centered cubic (FCC) and hexagonal close-packed (HCP)
structure. In both structures, the close-packed layers are shifted
with respect of each other, such that the centers of mass in one layer
are above the holes in the layer below. This leads to three different
layer positions labeled by A, B and C. The FCC crystal corresponds to
ABC stacking, while ABAB stacking is characteristic for the HCP
crystal \cite{Book:Kittel}. 

In the case of a densely packed crystal of cylinders, there is an
infinity of possible stackings positions, provided that the
close-packed planes are flat. However away from close packing, the
system 
will prefer a situation where the layers are stacked in an AAA
fashion. Other stackings are not stable and will relax to this
structure. 

If the cylinders are not confined to close-packed planes, the AAA
crystal phase could deform in a columnar phase. In this phase the
cylinders are organized in a hexagonal array of columns. However, there
is no long-range correlation between the longitudinal displacement of
different columns.

Another crystalline phase that is possibly relevant is the simple
cubic crystal, in which the orientation distribution of the cylinders
has the same symmetry as the lattice. Although this phase cannot be
close packed, it allows us prepare a structure that has the same
orientational symmetry as a cubatic phase. This is achieved by
aligning the cylinders at random with any of the three axes of the
crystal. 

\section{Simulation results}
\label{sec:result}

Constant NPT-MC simulations were performed on a system of 720
particles. Long runs were needed to collect good statistics (typically 
$10^5$ trial moves per particle excluding equilibration). On average
a volume change was attempted N times less frequently than the simple
particle moves. One in every ten trial rotations was a flip-move.

In our NPT simulations of the solid phase, we allowed the shape of the
simulation box to change to an arbitrary parallelepiped rather than
keeping it rectangular. Such a move was first used by Parrinello and Rahman
\cite{Parrinello:1980PRL,Parrinello:1981JAP,Parrinello:1982JCP} in MD
simulations and later by Najafabadi and Yip in MC simulations
\cite{Najafabadi:1983SM}. Allowing for shape changes is only 
useful in crystalline structures, because then the presence of the
crystal structure acts as a restoring force and will ensure that the
box shape cannot become extremely anisotropic. In a liquid phase,
extreme deformations could occur, because a natural restoring force
does not exists. 

\begin{center}
\begin{figure}[h]
\epsfig{figure=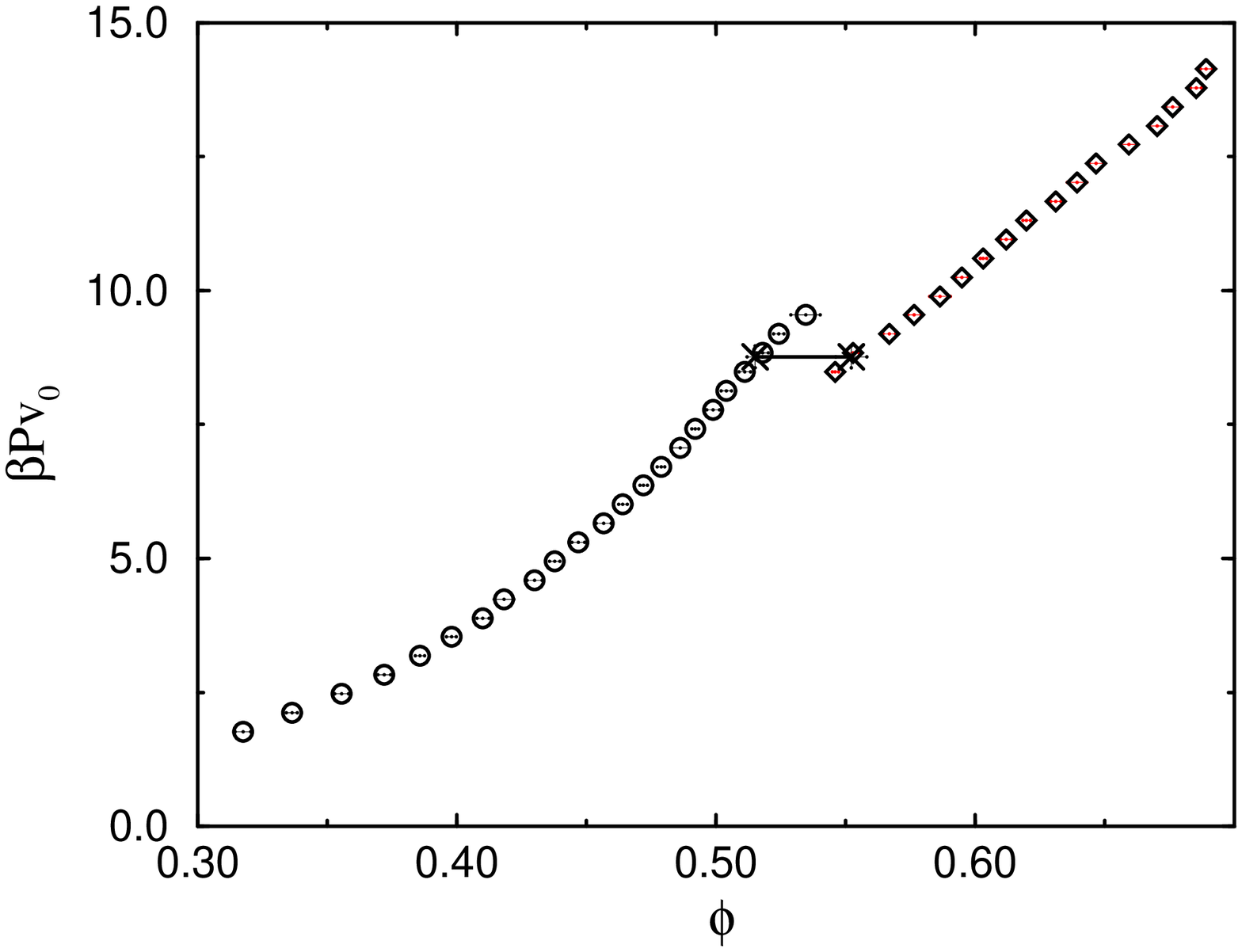,width=8.5cm}
\vspace{4cm}
\caption[a]{\label{fig:eq_of_state} R. Blaak, Journal of Chemical Physics}
\end{figure}
\end{center}

The resulting equation of state is shown in
Fig. \ref{fig:eq_of_state}, where we plotted the reduced pressure
$\beta P v$ versus the packing fraction $\phi$, where $v$ is the
volume of a single cylinder and $\beta = 1/(k_B T)$ the inverse
temperature. The two branches are obtained by compressing a
low-density liquid phase and expanding a high-density AAA-crystal. The
equation of state provides no indication that a cubatic phase, or any
other intermediate phase, exists.

If we start with a crystalline phase in which the hexagonal layers are
stacked in a different fashion, for instance ABC, the layers will
swiftly slide with respect to each other to form an
AAA-stacking. Occasionally, we find that the columns shift with
respect to each other, but this is probably a finite size effect,
rather than an indication of a columnar phase. 

We can also start from a simple cubic crystal in which the particles
are either regularly or randomly aligned with respect to the three
axes of the crystal. This phase is also not stable but tends to
transform into an the AAA-crystal. It is observed that this sometimes
leads to two AAA-crystals with different orientations. However, it is
likely that, given enough time, the grain boundary that has formed
will anneal out. 

The flip-move is extremely effective. Even for a packing fraction
$\phi= 0.7$ one out of every 1000 attempted flips is accepted. For
$\phi = 0.65$ this is already one out of every 100 and increases to
about 7\% near the transition to the isotropic phase. A simulation,
without using this flip-move, showed that at $\phi=0.65$ particles can
achieve a flip via a continuous rotation, but this occurs with a very
low probability, only once in a simulation of $10^5$ sweeps. Therefore
the introduction of the flip-move speeds up this slow `dynamical'
process enormously, leading to a faster convergence to equilibrium. 

\begin{center}
\begin{figure}[h]
\psfig{figure=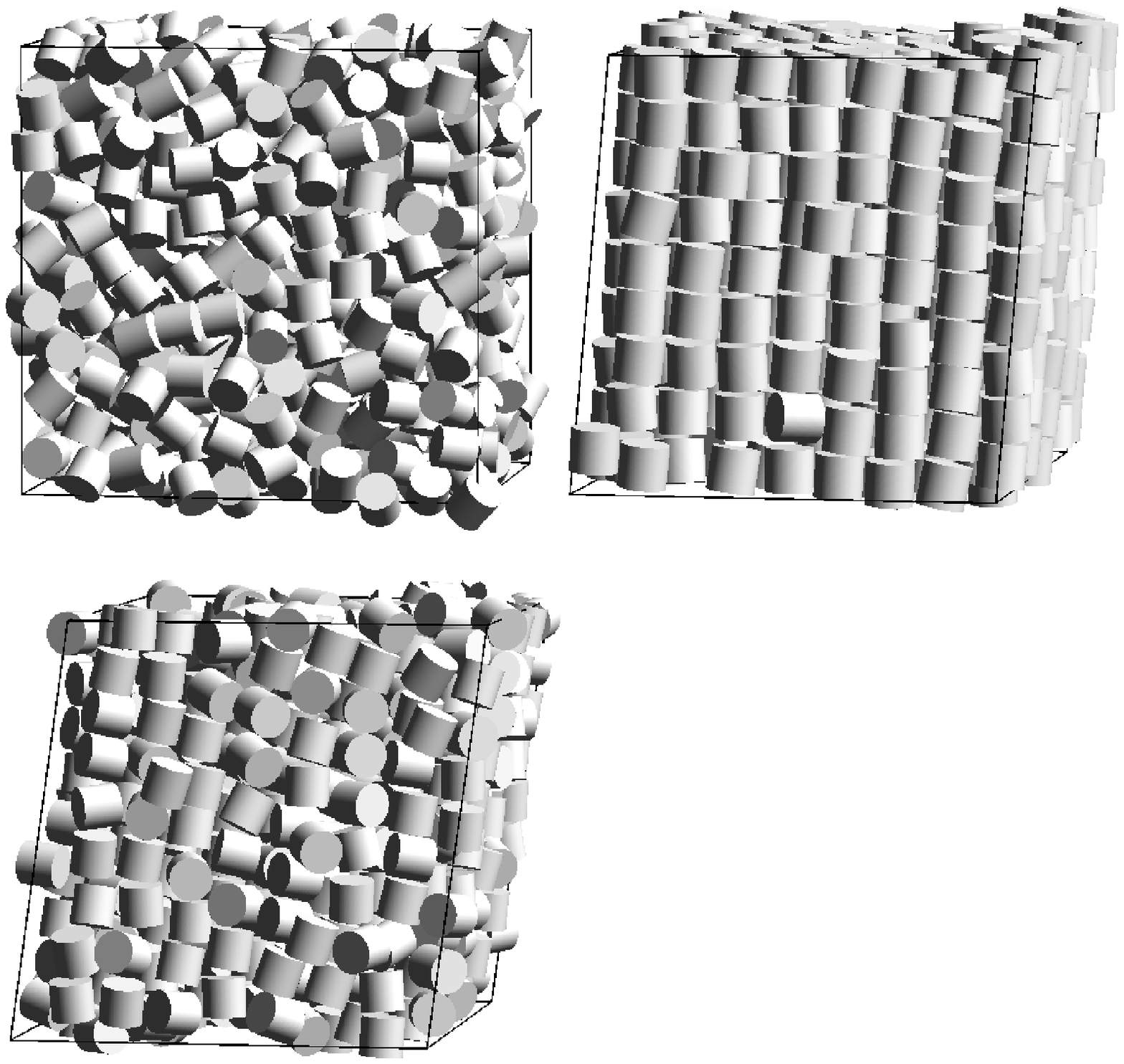,width=16cm,angle=0}
\vspace{4cm}
\caption[a]{\label{fig:snap-cyl} R. Blaak, Journal of Chemical Physics}
\end{figure}
\end{center}

In order to get an impression what these systems look like, three
snapshots are shown in Fig. \ref{fig:snap-cyl}. The first snapshot
shows a high-density isotropic phase. The second represents a
high-density, almost perfect AAA-crystal at a packing fraction
$\phi=0.684$, with one orientational defect. The third and most
interesting snapshot is obtained at the low-density side of the
crystalline branch. This is the place where, the cubatic phase would
be observed if it existed. The phase that we observe appears to be
neither isotropic nor perfectly crystalline. A stunning feature of
this phase is the fact that the shape of the simulation box changes
from a rectangular shape at high densities to a non-orthogonal shape
in the intermediate region. As the symmetry of the
structure differs from both the high density solid and the low density
liquid, we must classify it as a separate phase. Since the equation of
state shows no hint of a first-order phase transition, the novel phase 
is probably joined to the crystalline phase by either a continuous or
weakly first-order transition and is found at packing fractions
between $\phi \approx 0.55$ and $\phi \approx 0.65$.

\section{Symmetry of the intermediate phase}
\label{sec:symmetry}
In order to analyze the nature of the intermediate phase we
discuss the translational and orientational order separately. 
To characterize the positional order we use the bond orientational
order parameters $Q_l$ used by Steinhardt {\em et al.}
\cite{Steinhardt:1983PRB},   

\begin{equation}
\label{eq:sim-bond}
Q_l = \left\langle \left( \sum_m |C_{l,m}(\vec{r})|^2
\right)^{\mbox{$\frac{1}{2}$}} \right\rangle ,
\end{equation}
where we determine the polar angles of the vector $\vec{r}$
connecting neighboring particles. Only particles within a distance of 
twice the length of the particles are used. 

\begin{center}
\begin{figure}[h]
\epsfig{figure=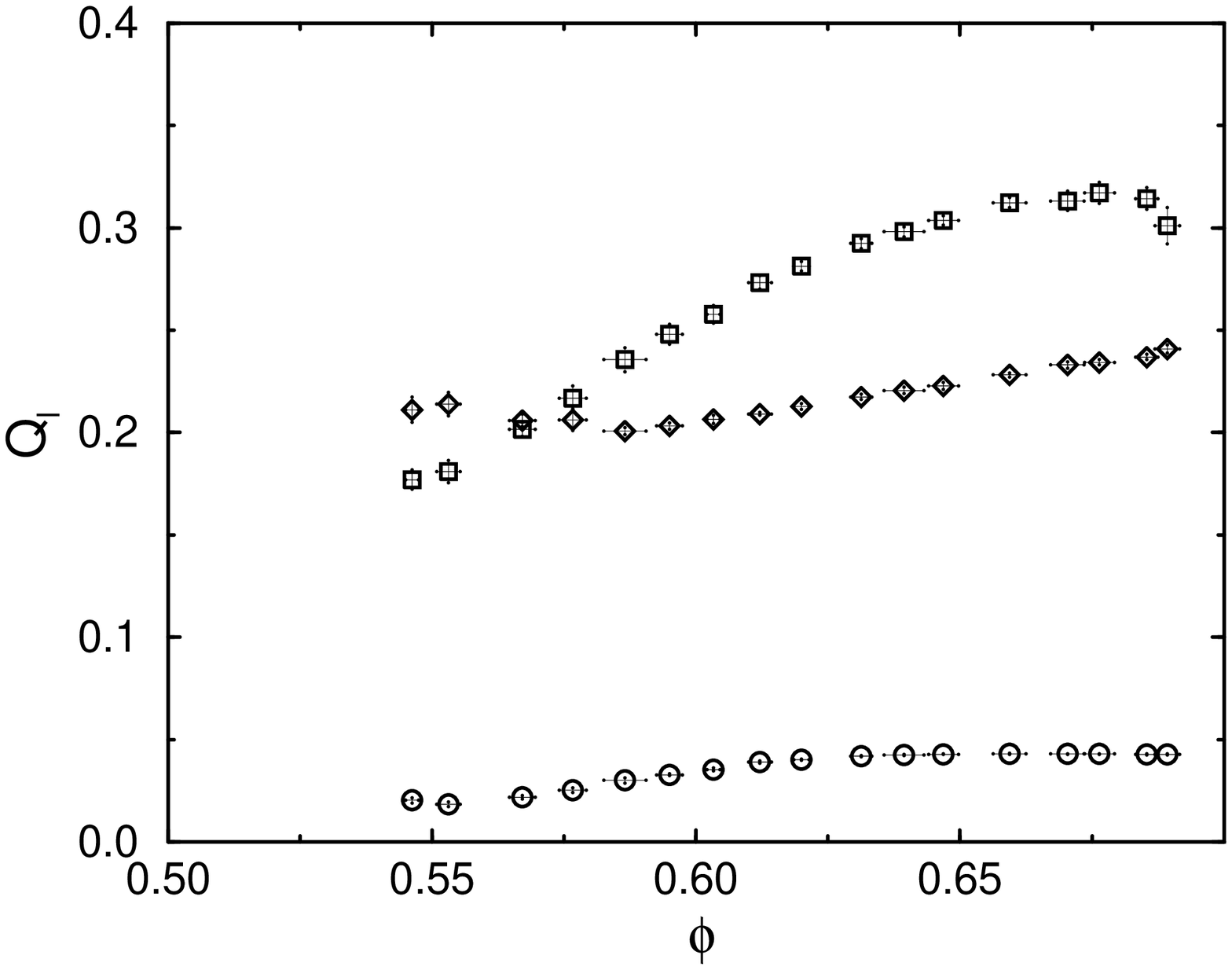,width=8.5cm}
\vspace{4cm}
\caption[a]{\label{fig:bond} R. Blaak, Journal of Chemical Physics}
\end{figure}
\end{center}

\begin{center}
\begin{figure}[h]
\psfig{figure=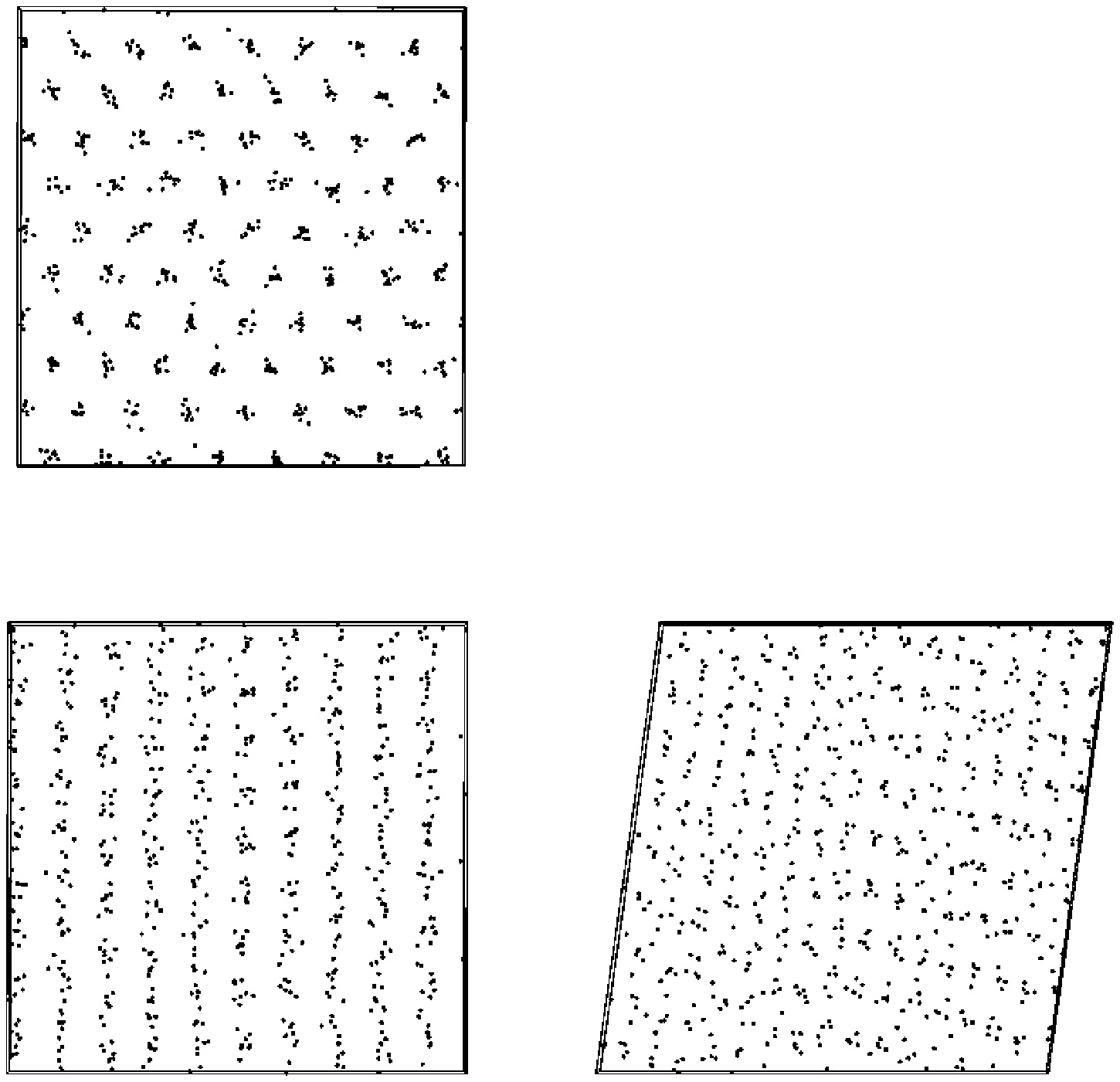,width=16cm,angle=0}
\vspace{4cm}
\caption[a]{\label{fig:snap-cyl-pro} R. Blaak, Journal of Chemical Physics}
\end{figure}
\end{center}

The bond orientational order parameters for $l=4$, 6 and 8 are shown
in Fig. \ref{fig:bond}. As can be seen from the figure, the behavior
of these order parameters does not indicate any sudden structural
phase transition. Figure \ref{fig:snap-cyl-pro} shows the projections
of the centers of mass of the particles close to melting  ($\phi =
0.567$). The top view shows clearly the hexagonal structure,
indicating the presence of columns. The side view shows that columns
lie within planes. 

\begin{center}
\begin{figure}[h]
\epsfig{figure=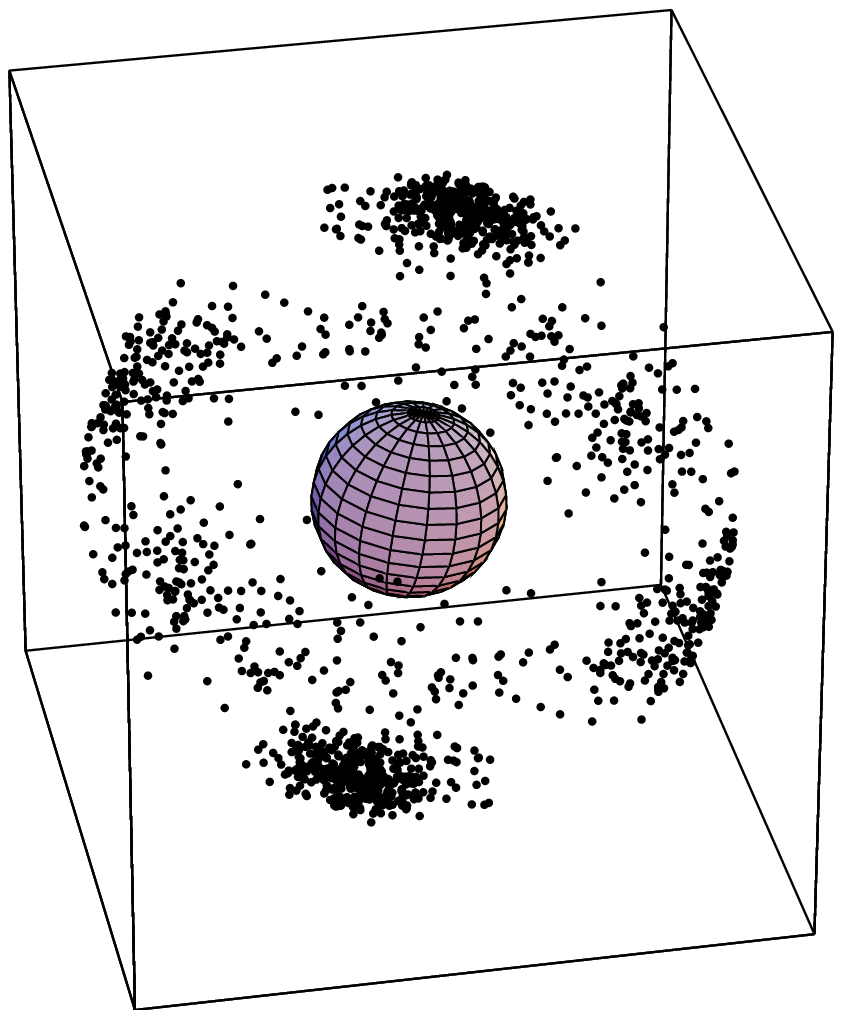,width=7cm}
\vspace{4cm}
\caption[a]{\label{fig:orientation} R. Blaak, Journal of Chemical Physics}
\end{figure}
\end{center}

We use the system at packing fraction $\phi = 0.567$ to illustrate the 
orientational order in the phase near the transition. In
Fig. \ref{fig:orientation} we plot all orientations of the particles
on a unit sphere. It is clear that particles are oriented either along
a main axis or in a plane perpendicular to that direction.  

In order to detect the orientational order we calculate the following
matrix 

\begin{equation}
\label{eq:sim-Q}
{\bf Q} = \frac{1}{N} \sum_{i=1}^N
\left( \mbox{$\frac{3}{2}$} \hat{u}_i 
\hat{u}_i - \mbox{$\frac{1}{2}$} {\bf I} \right),
\end{equation}
where $\hat{u}_i$ is the director of a particle and $N$ is the total
number of particles. The nematic order parameter $\bar{P_2}$ is defined as the
average value of the maximum eigenvalue of the matrix $Q$ and the
nematic direction as the corresponding eigenvector. Furthermore we
evaluate the order parameters $I_l$, defined by  

\begin{equation}
\label{eq:sim-cub}
I_l = \left\langle \left( \sum_m  C_{l,m} C_{l,m}^*
\right)^{\mbox{$\frac{1}{2}$}} \right\rangle.
\end{equation}
These order parameters are defined in terms of the modified spherical
harmonics $C_{l,m}$, and are frame independent. $I_2$ is related to
the nematic order parameter. But whereas the nematic order parameter
$\bar{P_2}$ measures uniaxial order only, $I_2$ measures the combination of
uniaxial and biaxial order. The order parameter $I_4$ is also able to measure
uniaxial order, but now via $P_4$, and biaxial order. In addition,
however, it is also able to measure cubatic order. It is therefore a
non-zero $I_4$ combined with the absence of nematic order, and thus a 
zero valued $I_2$, which would indicate the presence of cubatic order
in the system.  

\begin{center}
\begin{figure}[h]
\epsfig{figure=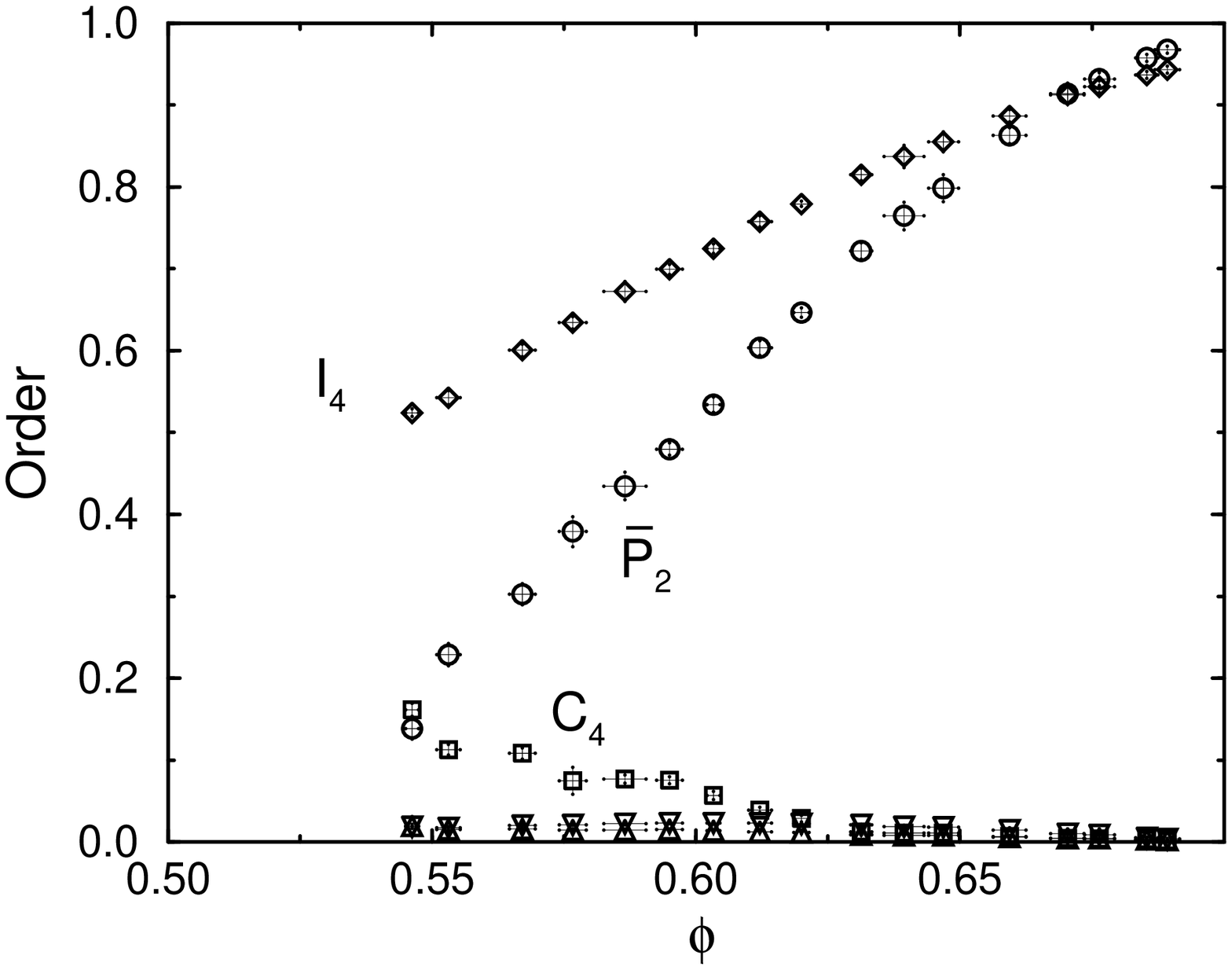,width=8.5cm}
\vspace{4cm}
\caption[a]{\label{fig:orient} R. Blaak, Journal of Chemical Physics}
\end{figure}
\end{center}

Both order parameters, $\bar{P_2}$ and $I_4$, are shown in Fig.
\ref{fig:orient}. The non-zero value of $\bar{P_2}$ on the complete
crystalline branch, indicates that there is a single preferred or
average direction in the system. It turns out that this direction is
along the direction of the columns of the crystal. 
For low density crystals, however, the value of $\bar{P_2}$ is small
and comparable to the nematic order close to the isotropic-nematic
phase transition. This is explained by the fact that
particles are either aligned along the columns or perpendicular to them.
For lower densities more particles will have there direction perpendicular to
the columns. It is not favorable to have a direction in between the
two extreme orientations. 

In order to detect the orientational order in the plane perpendicular
to the columns, we finally consider the average of the functions $\sin
l \varphi$, where the angles $\varphi$ of the directors of the
particles are measured in the plane perpendicular to 
the average direction obtained from (\ref{eq:sim-Q})
\begin{equation}
\label{eq:sim-C}
C_l = \langle | C_{l,l}| \rangle.
\end{equation}
These order parameters are also shown in Fig. \ref{fig:orient}, and
indicate that there is a small, 4-fold symmetry present. This is
caused by the preference of particles in a layer to be either
parallel or perpendicular. The 4-fold symmetry is, however, in
conflict with the 6-fold symmetry of the crystal. These conflicting
symmetries are probably the reason of the non-orthogonal unit cell.

\begin{center}
\begin{figure}[h]
\epsfig{figure=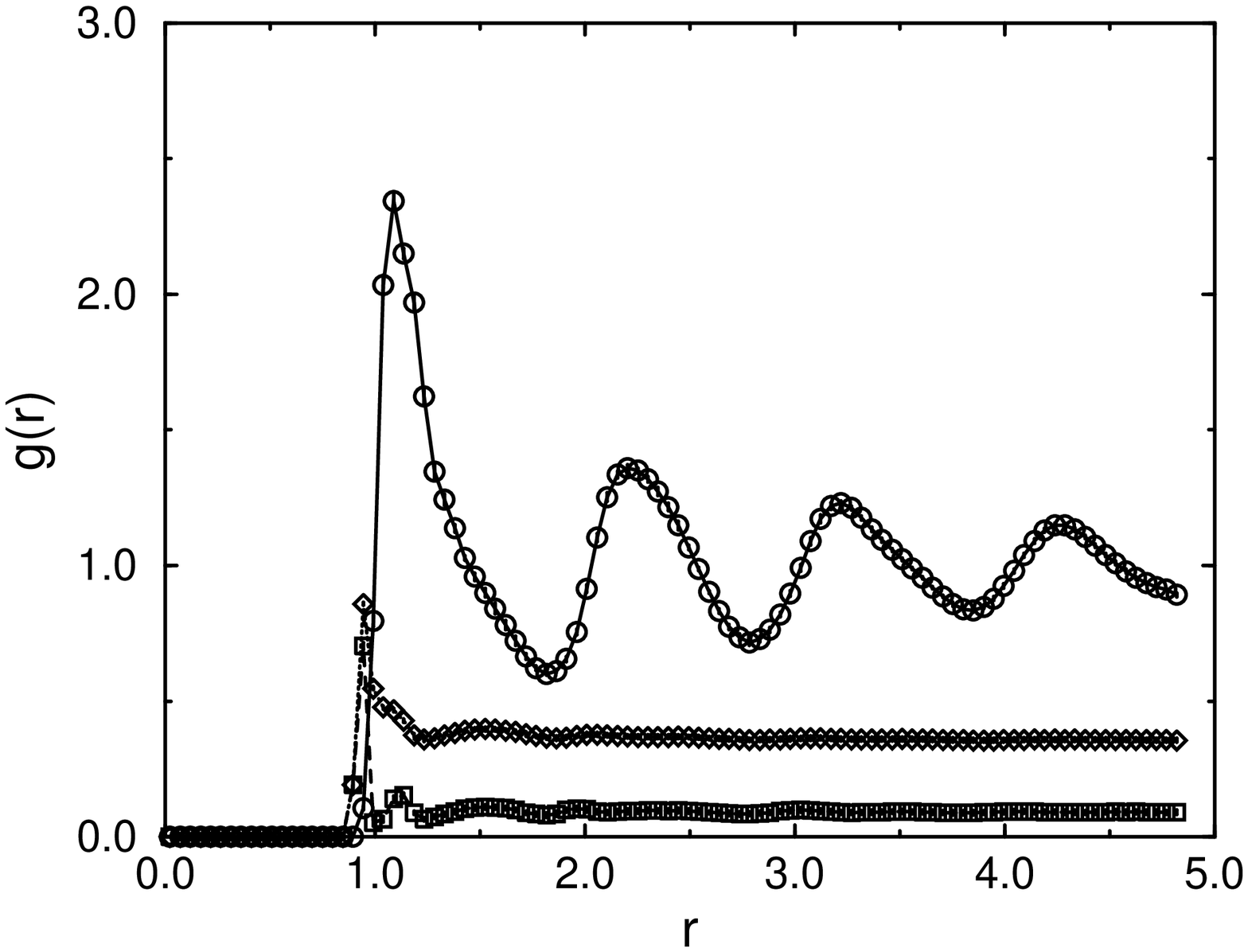,width=8.5cm}
\vspace{4cm}
\caption[a]{\label{fig:radial} R. Blaak, Journal of Chemical Physics}
\end{figure}
\end{center}

In Fig. \ref{fig:radial} we show the radial distribution function
$g(r)$ and the orientational correlation functions $g_l(r)$ for $l=2$
and 4, that are defined by 

\begin{equation}
\label{eq:radial}
g_l(r) = \langle P_l(\hat{u}(0).\hat{u}(r))\rangle,
\end{equation}
where $P_l$ is the $l$th Legendre polynomial and $\hat{u}(r)$ is a
unit vector along the axis of a particle at distance r from the
reference particle. These correlation
functions are a measure of the range of the orientational order. As
can be seen from Fig \ref{fig:radial}, $g_2(r)$ and $g_4(r)$ are finite
throughout the complete simulation box. Although the correlation
function $g_4(r)$ is measuring a combination of nematic-like and
cubatic-like order, the magnitude of $g_4(r)$ compared to that of
$g_2(r)$ indicates that the cubatic-like order is dominant. This is
different from a normal nematic phase where $g_2(r)$ is larger that
$g_4(r)$. 

\begin{center}
\begin{figure}[h]
\epsfig{figure=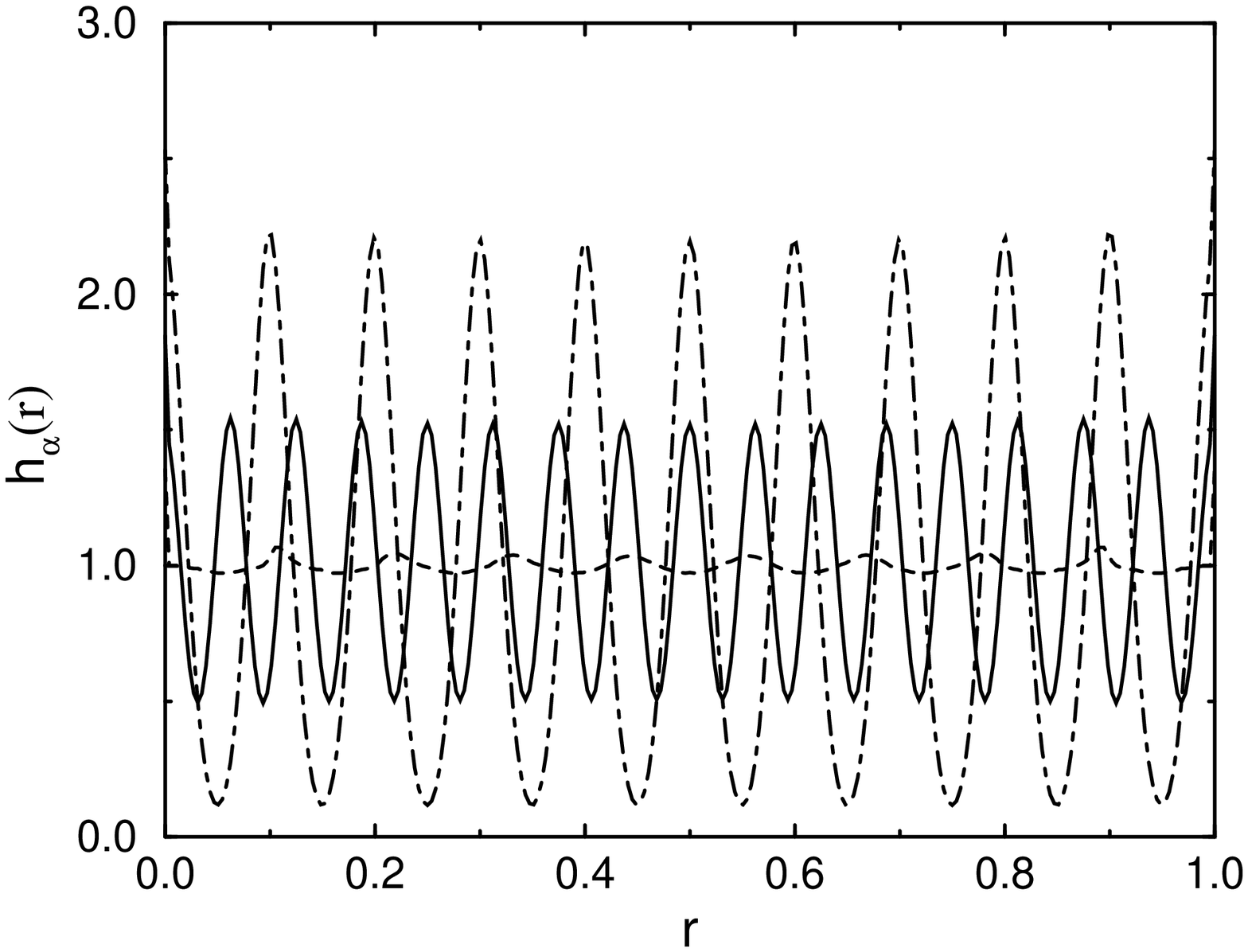,width=8.5cm}
\vspace{4cm}
\caption[a]{\label{fig:density} R. Blaak, Journal of Chemical Physics}
\end{figure}
\end{center}

Finally we show in Fig. \ref{fig:density} the density correlation
functions $h_{\alpha}(r)$ defined by

\begin{equation}
h_{\alpha}(r) = \langle \rho(0) \rho(r) \rangle /\rho^2,
\end{equation}
where the distance $r$ is measured along the $\alpha$ direction and in
units of the simulation box length. The functions $h_x$ and $h_y$ show
distinct features, as could be expected from the projections shown in
Fig. \ref{fig:snap-cyl-pro}, and the number of peaks correspond to the
number of layers present in that direction. In the $z$ direction a
weak modulation 
 is observed, although the peaks of the function $h_z$ are less
pronounced. This can partly be understood if one realizes that the
fluctuations in the positions along the columns can be of the order of
20\% of the effective size of the particles in that direction. These
fluctuations are even enhanced by the fact that particles are oriented
along and perpendicular to the columns, but the effective size in both
cases is different. For the system sizes that we studied, the density
modulation in the $z$ direction persists over the entire simulation
box. However, we cannot rule out the possibility that the modulation
will decay in even larger systems. If this were the case, the system
would, in fact, be in a columnar phase.

In summary, in our simulations we find evidence for an
intermediate phase with 3D-crystalline structure in
which hexagonal layers are stacked in an AAA fashion. The orientations
of the particles are coupled to the crystal axes and are either along the
columns or within the layers, in which case they have a slight 4-fold 
symmetry. Neither the equation of state, nor the order parameters show
any indication that there is a first order phase transition along the
crystal branch to the perfect, aligned crystal. The novel phase is
found at packing fractions between $\phi \approx 0.55$ and $\phi
\approx 0.65$, although this upper boundary is somewhat arbitrary as
we found no clear phase transition. Above this boundary, however, less
than 1\% of the particles is misaligned with the crystal. 

\section{Free energy calculation}
\label{sec:free}

In order to determine the coexistence between the isotropic liquid and
the ordered phase we have to find points on the equation of state with
equal pressure and chemical potential. To this end we need to determine
the free energy. By performing a thermodynamic integration along the
equation of state we can evaluate the free energies up to a constant.
This integration is only allowed if there is no first order transition
along the integration path, which in our case is true because all
evidence indicates that the transition on the crystal branch is continuous.

In order to proceed we need to determine one reference point on either
branch of the equation of state. In the isotropic phase this is easy
because we can use the ideal gas as a reference state

\begin{equation}
\label{eq:free-int-gen}
F(\rho) = F_{id}(\rho) + \int_0^{\rho} d \rho'\frac{P(\rho') -
\rho'}{{\rho'}^2}.
\end{equation}
For the crystal phase we use as a reference system an Einstein crystal
with the same structure \cite{Frenkel:1984JCP}, which in our case is a
perfectly aligned AAA-crystal. We connect the two systems by a one
parameter Hamiltonian which consists of two parts, the coupling of the
particles to their equilibrium lattice positions and the alignment
for the orientations

\begin{equation}
\label{eq:ham}
H_{\lambda} = \lambda \sum_i (\vec{r}_i - \vec{r}_i^0)^2 + \lambda
\sum_i \sin^2(\theta_i),
\end{equation}
where $\lambda$ is the coupling parameter, $\vec{r}_i^0$ are the
lattice positions, $\vec{r}_i$ the positions of the particles and 
$\theta_i$ are respectively the orientation of the particles with
respect to the preferred direction. By slowly increasing the
value of $\lambda$ the system will order according to the imposed field. 
We used here the same coupling parameter for both fields, but one can
also use different values to do the orientational and positional
ordering separately. 

The free system with $\lambda=0$ can be related to the Einstein
crystal for which $\lambda \gg 1$ by

\begin{equation}
\label{eq:free-int}
\frac{\beta F(\rho)}{N} = \frac{\beta F_{ein}(\lambda_{max})}{N} -
\int_0^{\lambda_{max}} d \lambda <\delta r^2>_{\lambda} - 
 \int_0^{\lambda_{max}} d \lambda <\sin^2\theta>_{\lambda} - 
\frac{\log V}{N},
\end{equation}
where $<\delta r^2>_{\lambda}$ is the mean square displacement and
$<\sin^2\theta>_{\lambda}$ the average sine squared of the angle
between the directors of the particles and the direction of the
alignment field. The last term corrects for the fact that the center of
mass during this simulation needs to be fixed. The value of
$\lambda_{max}$ is chosen such that a system using this Hamiltonian
only will not cause any overlaps in the system. For that limit we can
derive the value of the free energy of the Einstein crystal 
\cite{Polson:1998Un}

\begin{equation}
\label{eq:free-ein}
\beta F_{ein}(\lambda) = -\mbox{$\frac{1}{2}$} \log N -
\mbox{$\frac{3}{2}$} (N-1)
\log(\frac{\pi}{\beta \lambda})  - N \log(\frac{2 \pi}{\beta \lambda}).
\end{equation}
In the case that overlaps do occur, it is possible to correct for
this \cite{Frenkel:1984JCP}. By simulating the system for different values of
$\lambda$ the integrals in (\ref{eq:free-int}) can be evaluated
numerically. In order to minimize the error we used the Gauss-Legendre
quadrature. 

The coexistence is obtained is at packing fraction ${\phi=0.515 \pm
0.003}$ for the isotropic and  ${\phi=0.552 \pm 0.006}$ for the
crystalline structure, at a pressure ${\beta P v_0 = 8.76 \pm 0.08}$
(Fig. \ref{fig:eq_of_state}). 

\section{Discussion}
\label{sec:discussion}
Our study of the hard cylinder system was inspired by the possibility
that it might exhibit a cubatic phase. We have presented Monte Carlo
simulation for cylinders with an aspect ratio $L/D=0.9$. In the
standard NPT-simulations we introduced the flip-move, which rotates a
particle directly to a perpendicular orientation. Furthermore we
allowed the box shape to change to an arbitrary parallelepiped to
facilitate equilibration of the crystalline phase. 

At high pressures the system is found in an AAA-crystal with particles
aligned along the direction perpendicular to the layers. By lowering
the pressure, particles reorient, deform the perfect crystal and form
an intermediate phase in which the 6-fold symmetry of the crystal is
slightly distorted and only very weak layering can be observed.
The orientations are either along the columns or perpendicular to
them. Towards the phase transition to the isotropic liquid, a weak, but
significant, 4-fold orientational order in plane develops. There is no
indication that this structural phase transition is first order. The
phase coexistence between the isotropic and crystal phase was
determined at packing fractions $\phi=0.515$ in the isotropic phase
and  $\phi=0.552$ for the crystal phase. 

Our simulations did not reveal a true cubatic phase. In fact, a
theoretical analysis of this system \cite{Blaak:1997PhD} suggests that 
the isotropic to cubatic phase transition occurs at a density beyond
that of the isotropic to nematic phase transition. This theory
predicts that the isotropic to cubatic transition would take place at
a packing fraction $\phi \approx 0.80$. However, at this density
positional order cannot be neglected and should be incorporated in the
theoretical description. 

The present results suggest that it is an oversimplification to
represent stacks of cut-spheres by monodisperse hard cylinders. As can
be seen in Fig. \ref{fig:cubatic}, the stacks are polydisperse in
length and shape. It might be possible that a cubatic 
phase is stabilized by introducing polydispersity to the
system. However, such a study is outside the scope of the present
paper.  

\section*{Acknowledgments}
We thank Martin Bates for a critical reading of the manuscript. 
The work of the FOM Institute is part of the research
program of FOM and is made possible by financial support from the 
Netherlands Organization for Scientific Research (NWO).

\newpage
\begin{center}
{\large FIGURE CAPTIONS}
\end{center}

\begin{enumerate}

\item
Snapshot of the cubatic phase for cut-spheres.

\item
The excluded volume of two identical cylinders as function
of the mutual angle $\gamma$ for different aspect ratios $L/D=$ 0 and
$\infty$ (solid), 1 (dashed), 2 (long dashed) and 10
(dotted-dashed). The excluded volume is normalized to unity for
perpendicular orientations.

\item
The ratio of excluded volume for perpendicular and parallel
orientations as a function of the aspect ratio of the cylinders. The
minimum is found for $L/D=\sqrt{\pi}/2$.

\item
Top view of a hexagonal structure for cylinders with aspect
ratio of order unity. The left figure shows the aligned case and the
middle figure the case when the particle is perpendicular to a layer.
The right figure shows an intermediate orientation causing overlap.

\item
The equation of state of cylinders with $L/D = 0.9$ for the
isotropic (\symcircle) and crystal (\symdiamond) phase. The reduced
pressure $\beta P V_0$, where $v_0$ is the volume of a single
particle, is plotted versus the packing fraction $\phi$. The point of
coexistence between the isotropic and crystal branch are connected by
the solid line.

\item
Three snap shots from simulations, corresponding to isotropic
(top left), high density crystal (top right) and ordered phase near
the transition (bottom). The packing fractions for these phases are
respectively 0.478, 0.684 and 0.567.

\item
The bond orientational order parameters $Q_4$(\symcircle),
$Q_6$(\symdiamond) and $Q_8$(\symsquare) as function of the packing
fraction in the ordered phase.

\item
The projections of a configuration at packing fraction
$\phi=0.567$. The top view (top) shows the hexagonal pattern of the
columns which can be seen from the side (under left). In the front
view (right under) planes are more difficult to observe.

\item
The orientational distribution for a configuration at $\phi=0.567$.

\item
The orientational order parameters as function of the packing
fraction in the ordered phase. The nematic (\symcircle), cubatic
(\symdiamond) and  $C_4$ (\symsquare) order parameters are non-zero,
$C_2$ (\symtriangleu) and $C_6$ (\symtriangled) are negligible small.

\item
The radial distribution function $g(r)$ (\symcircle) and the
orientational correlation functions $g_2(r)$ (\symsquare) and $g_4(r)$
(\symdiamond) as a function of the distance $r$ measured in units of
the diameter of the cylinders at $\phi=0.567$.

\item
The correlation functions $h_{\alpha}(r)$ as function of the distance
along the axis and measured in units of the box length at
$\phi=0.567$. $h_x$ (solid) and $h_y$ (dotted-dashed) clearly
indicate the presence of layers in the $x$ respectively $y$
direction. Although $h_z$ (dashed) has less pronounced peaks,
also here layering is visible.

\end{enumerate}

\newpage

\newpage

\newpage

\newpage

\newpage

\newpage

\newpage

\newpage

\newpage

\newpage

\newpage

\end{document}